\documentclass[final]{aipproc}

\layoutstyle{8x11single}

\begin{document}

\title{Single-photon generation and simultaneous observation of wave and particle properties}

\author{Thomas Aichele, Ulrike Herzog, Matthias Scholz and Oliver Benson}{
  address={Humboldt-Universit\"at zu Berlin, Nano Optics, Hausvogteiplatz 5-7, 10117 Berlin, Germany},
  email={thomas.aichele@physik.hu-berlin.de}
  homepage={http://nano.physik.hu-berlin.de/}
}

\begin{abstract}
We describe an experiment that generates single photons on demand and measures properties accounted
to both particle- and wave-like features of light. The measurement is performed by exploiting data
that are sampled simultaneously in a single experimental run.
\end{abstract}

\maketitle

\section{Introduction}

The wave-particle duality of matter lies at the heart of quantum mechanics. With respect to light,
the wave-like behavior is perceived as being classical and the particle aspect as being
nonclassical, while for massive microscopic objects, like neutrons and atoms, the opposite holds.
The occurrence of an interference pattern is a manifestation of the wave-nature of matter. In the
frame of classical wave optics, standard first-order interference is explained by the superposition
of the electric field strengths when two coherent partial beams merge. This superposition can be
constructive or destructive and depends sensitively on the phase difference between the partial
beams, thus giving rise to a spatial or temporal interference pattern when the latter varies.

Already in 1909, soon after the introduction of the concept of the photon, it was observed
experimentally that there is no deviation from the classically predicted interference pattern if a
double-slit interference experiment is performed with very weak light, even if the intensity is so
small that on average only a single photon is present inside the apparatus \cite{Taylor}. Later
this observation was accounted for theoretically by quantum mechanics and was confirmed by more
precise experiments \cite{Janossy,ReynoldsSpartalianScarl}. There exists an exact correspondence
between the interference of the quantum probability amplitudes for each single photon to travel
along either path in an interferometer on the one hand and the interference of the classical field
strengths in the different paths on the other hand. Therefore, the outcome of any first-order
interference experiment can be obtained by describing light as a classical electromagnetic wave,
independent of the statistical distribution of the incident photons. Both wave- and particle-aspect
can be observed in a single experiment when with very small light intensities from a classical
source the double-slit interference pattern is gradually built up by registering more and more
spots on the screen. A small non-vanishing probability remains that such a spot is not caused by a
single photon, but by two photons arriving at the same time, though. According to the principle of
complementarity, it is impossible to simultaneously observe interference and to detect which path
each photon travelled in the interferometer.

\begin{figure}
\includegraphics[width=0.8\textwidth]{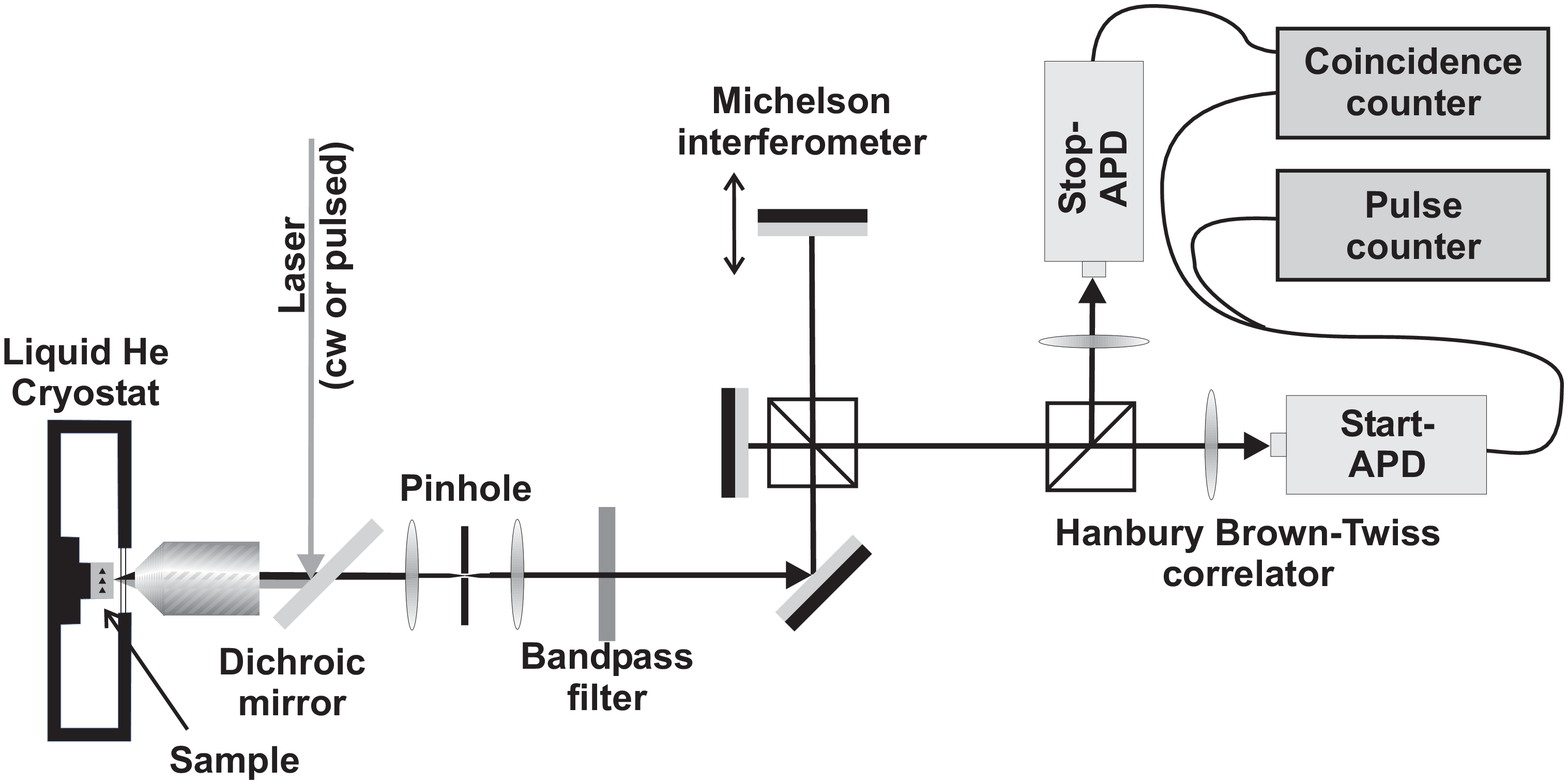}
\caption{\label{Setup}Basic experimental setup.}
\end{figure}

In this paper we do not raise the question of complementarity, but report an experiment at the
exact single-photon level to detect wave- and particle-like properties of light in a single run.
For this purpose we use the fact that the nonclassical, or corpuscular, aspect of light can be
revealed by observing intensity correlations using a setup that was originally invented by Hanbury
Brown and Twiss for determining the diameter of stars \cite{HBT1,HBT2}. In this scheme, intensity
correlations between two partial beams are measured dependent on the path difference or the mutual
time delay, respectively. These correlations correspond to delayed coincidences between the clicks
of two detectors, each detecting the photons in one of the partial beams. Photon antibunching
occurs when the coincidence probability increases with growing delay time. This effect cannot be
explained by classical wave theory and is a clear indication for the particle-like nature of light
(see e. g. \cite{WallsMilburn}). Kimble et al. were the first to detect photon antibunching, using
the light from atomic resonance fluorescence \cite{KimbleMandel}. Later, Grangier et al.
\cite{Grangier} performed a series of experiments with single photons from atomic decays. In a
first step, they showed the single-photon character of the atomic emission by observing the
corresponding antibunched behavior of the intensity correlation function. In a second step, they
inserted the photons into a Mach-Zehnder interferometer, and observed an interference pattern with
varying path difference, a feature that displays the wave nature of light. Braig et al.
\cite{KurtsieferWaveParticle} implemented a similar experiment using a diamond defect center as
emitter, where they observed single-photon statistics after detecting interference in a Michelson
interferometer. We report an experiment where both parts are combined in a single step, similar to
the work of H\"off\-ges et al. \cite{H"offges} for simultaneously performing heterodyne and photon
correlation measurements in the resonance fluorescence of a single ion. Our light source consists
of single quantum dots, capable of producing single photons on demand.

\section{Experimental techniques and theoretical background}

\subsection{The single-photon source}

Single quantum dots are nanometer-scaled semiconductor islands that show discrete energy levels,
according to the spatial confinement of the charges. Due to this behavior, they are often referred
to as artificial atoms. An overview of the quantum dots topic can be found in \cite{Bimberg}. When
excited optically or electronically, a quantum dot acts as a trap for pairs of electrons and holes
that form a bound state, called exciton. This quasi-particle emits a single photon at radiative
decay, similar to an optical atomic transition. Single quantum dots have turned out to be stable
and efficient single photon sources that cover the whole visible and large parts of the infrared
spectrum \cite{InP,SantoriIndistinguishable,CdSe,Michler}.

Our systems of choice are InP quantum dots embedded in a GaInP matrix. They emit around 700 nm,
where Si-based photodetectors reach their maximum detection efficiency.  Our basic experimental
setup is displayed in fig.~\ref{Setup}. The quantum dots have to be cooled by a liquid Helium
cryostat below 50~K (typically 10 K). Cooling is necessary because, at higher temperatures,
thermionic emission enables the charge carriers to escape the quantum dot before recombining
radiatively. Moreover, the coupling to phonons gives rise to spectral linewidth broadening which
deteriorates the filtering of individual optical transitions, as described below.

A microscope objective is used to focus the excitation laser onto the sample and also to collect
the photoluminescence light. The objective is chosen to have the highest possible numerical
aperture (NA=0.75) in order to increase the ratio of collected to overall-emitted light. The
numerical aperture is limited due to a minimally required working distance, given by the dimensions
of the geometry of the cryostat. The  density of the quantum dots is small enough to enable the
spatial resolution of individual quantum dots on the sample. The excitation lasers are optionally a
frequency doubled continuous Nd:YVO$_4$-laser at 532~nm wavelength or a frequency doubled pulsed
Ti:sapphire laser at a wavelength of 800 nm, a pulse width of 200~fs (FWHM) and a repetition time
of 13.2 ns.

\begin{figure}
\includegraphics[width=0.8\textwidth]{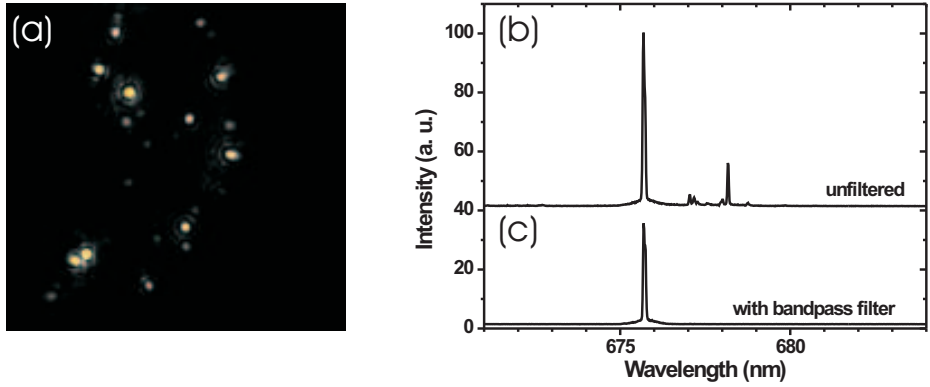}
\caption{\label{SingleQDs}(a) Microscope image of a \mbox{20 $\mu$m$\times$20 $\mu$m.} area on the
quantum dot sample. (b) Spectrum of a single quantum dot. (c) Spectrum behind a narrow bandpass
filter sparing only the line from the exciton decay.}
\end{figure}

Spatial and spectral filtering with a pinhole and a narrow bandpass filter (bandwidth 1.2 nm FWHM),
respectively, leaves only an individual quantum dot transition: Figure \ref{SingleQDs}(a) shows a
microscopic image of such a set of quantum dots, taken through the bandpass filter. The size of the
quantum dot images is determined by the resolution of the microscope. After positioning the sample
under the pinhole of the spatial filter, it is possible to select a particular quantum dot with the
help of a step motor. A typical quantum dot spectrum can be seen in fig. \ref{SingleQDs}(b). There
are mainly two quantum dot transitions present, resulting from the decay of a single exciton (one
electron-hole pair) and a bi-exciton (two electron-hole pairs, respectively. Behind the bandpass
filter, only photons from one transition of a single quantum dot is transmitted (fig.
\ref{SingleQDs}(c)).

\subsection{The Michelson interferometer}

Today, Michelson interferometry \cite{Michelson,MandelWolf} is a standard tool for single-photon
high resolution spectroscopy \cite{SantoriIndistinguishable,InPMichelson}. Our Michelson
interferometer for observing the wave-like character of the emitted light consists of an arm with a
fixed retro-reflecting mirror and another arm with a retro-reflector mounted on a piezo translator
with a range of 100 $\mu$m and a mechanical translator with a 20 mm range. The mechanical
translator was used to roughly align the interferometer arms.  The light field is split and merged
at a 50:50 beam splitter after a possible delay in one of the interferometer arms. The intensity of
the merged field varies with the path difference between the interferometer arms as long as it is
smaller than the coherence length of the light. In detail, the time-averaged intensity is given by
\begin{eqnarray}\label{EqnInt}
    \langle I(t,\tau)\rangle &=& 2\langle |E(t)|^2\rangle + 2\mathrm{Re}\langle
    E^*(t)E(t+\tau)\rangle \nonumber\\
    &=& 2\langle |E(t)|^2\rangle + 2 \langle|E(t)E(t+\tau)|\rangle\cos(\omega_o\tau), \nonumber\\
    &&
\end{eqnarray}
where $E(t)$ is the complex electric field and $\tau c$ is the path difference between the two
interferometer arms. The second equality sign holds for a quasi-monochromatic field
$E(t)=|E(t)|\exp(i\omega_ot)$ centered around the frequency $\omega_0/2\pi$. The interference
pattern is a sine-like variation around an average intensity. The contrast of the interference
fringes is characterized by the visibility
\begin{equation}
v=\frac{I_{\mathrm{max}}-I_{\mathrm{min}}}{I_{\mathrm{max}}+I_{\mathrm{min}}},
\end{equation}
with $I_{\mathrm{max}}$ and $I_{\mathrm{min}}$ being the maximum and minimum values of the
intensity, respectively. The visibility ranges from 0 (no interference at a constant intensity) to
1 (maximum interference, where the intensity can completely vanish). It can be also shown
\cite{MandelWolf} that the visibility is the absolute value of the Fourier transform of the optical
spectrum $S(\omega)$. For example a Lorentzian spectral line
$S(\omega)\propto1/\left((\omega-\omega_0)^2+\gamma^2\right)$ with a linewidth $\gamma$ has an
exponentially shaped visibility $v(\tau)\propto\exp(-\gamma\tau)$.

\subsection{The Hanbury Brown-Twiss correlator}

\begin{figure}
\includegraphics[width=\textwidth]{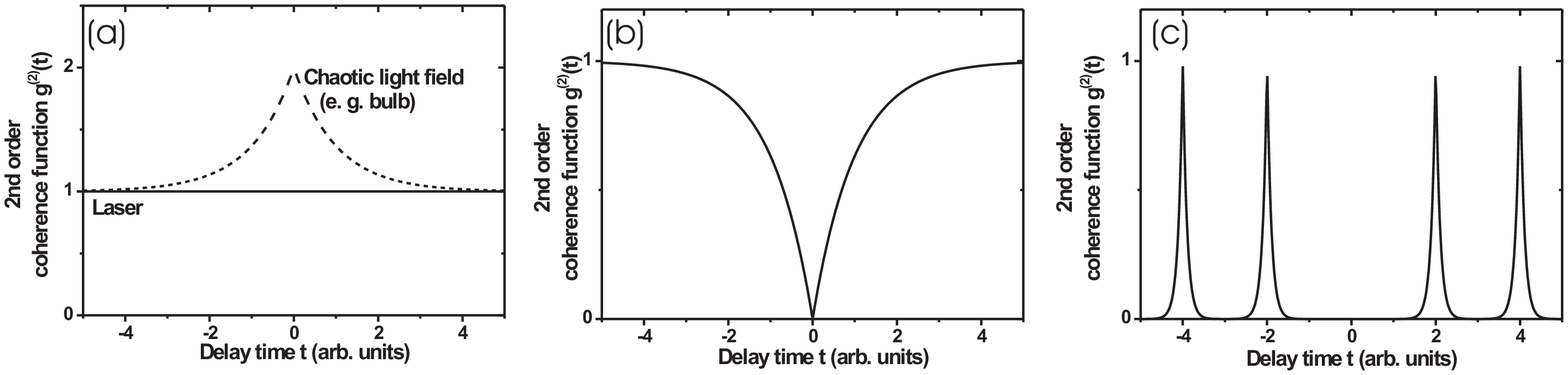}
\caption{\label{CorrFunc} Normalized correlation function $g^{(2)}(\tau)$ for different light
fields: (a) Classical field from an ideal laser (solid line) and from a thermal light source
(dashed line), (b) continuously pumped single quantum emitter and (c) single quantum emitter with
pulsed excitation, where every excitation pulse produces only one photon.}
\end{figure}

In order to give evidence of the particle aspect of the emitted light, we use a Hanbury Brown-Twiss
correlator which is a common tool to characterize the single-photon emission from quantum dots
\cite{InP,SantoriIndistinguishable,CdSe,Michler}. Our correlator consists of a 50:50 beam splitter
that divides the incident light field and of two single-photon detectors (avalanche photodetectors,
APDs). A time-interval counter collects the time differences between the start and the stop
detection events triggered by the two detectors, each from a different output beam. We used a
time-to-amplitude converter that collects the coincidences into 37-ps time bins. The time
resolution of the whole system was measured to be 800 ps.

After repeating the start-stop measurement many times, the distribution of the measured time
intervals is proportional\footnote{In fact, this proportionality only holds for times that are much
smaller than the average time distance $\bar t$ between adjacent detection events. In our
experiments the photon count rate is on the order of $50\times10^3$ counts per second. This gives
an average time distance of $\bar t \approx20$ $\mu$s which is much larger than the time scale of a
few nanoseconds for observing anti-correlation effects.} to the intensity correlation function,
given by the quantum-mechanical two-time expectation value $\langle \hat a^\dagger(0)\hat
a^\dagger(\tau)\hat a(\tau)\hat
    a(0)\rangle$, where $\hat a^\dagger(t)$ and $\hat a(t)$ are the photon
    creation and annihilation operators,
respectively, that act at time $t$. The normalized second-order coherence function is defined as
\begin{equation}
    g^{(2)}(\tau)=\frac{\left\langle \hat a^\dagger(0)\hat a^\dagger(\tau)\hat a(\tau)\hat
    a(0)\right\rangle}{\left\langle\hat a^\dagger(0)\hat a(0)\right\rangle^2} \, .
\label{g}
\end{equation}
For a stationary light field, the correlations between the photon detection events vanish after a
sufficiently large time delay and the coincidences become completely random. Therefore in the
stationary case $g^{(2)}(\tau)\rightarrow 1$ for $\tau \rightarrow \infty$. If $g^{(2)}(\tau)= 1$
for all values of $\tau$, the field exhibits a Poissonian photon statistics, where the variance of
the photon number is equal to its mean value when counted over intervals of arbitrary length. It
can be shown that for any classical light field  $g^{(2)}(\tau)\geq1$ and $g^{(2)}(0)\geq
g^{(2)}(\tau)$ \cite{WallsMilburn}. However, for nonclassical states of light, the correlation
function can have a very different shape with $g^{(2)}(\tau)<1$ at certain values of $\tau$, and
$g^{(2)}(0)< g^{(2)}(\tau)$. The latter inequality describes photon antibunching and is a clear
indication of the nonclassical, or corpuscular, character of the light. A value $g^{(2)}(0)<1$
implies that the appearance of two adjacent photons is less likely than in a random Poissonian
distribution with the same mean photon number. We find from Eq. \eqref{g} that $g^{(2)}(0) = 1-
1/n$ for a photon-number state $|n\rangle$, containing exactly $n$ photons. It is obvious that
$g^{(2)}(0) = 0$ for a single-photon state $|n=1\rangle$, because here, it is impossible to detect
two photons at the same time. When $g^{(2)}(0)$ is observed to be smaller than $0.5$, this is an
evidence for single-photon emission. Fig. \ref{CorrFunc} gives examples for the theoretically
predicted behavior of $g^{(2)}(\tau)$ for four different light fields.

\section{Measurement results and discussion}

\begin{figure}
\includegraphics[width=\textwidth]{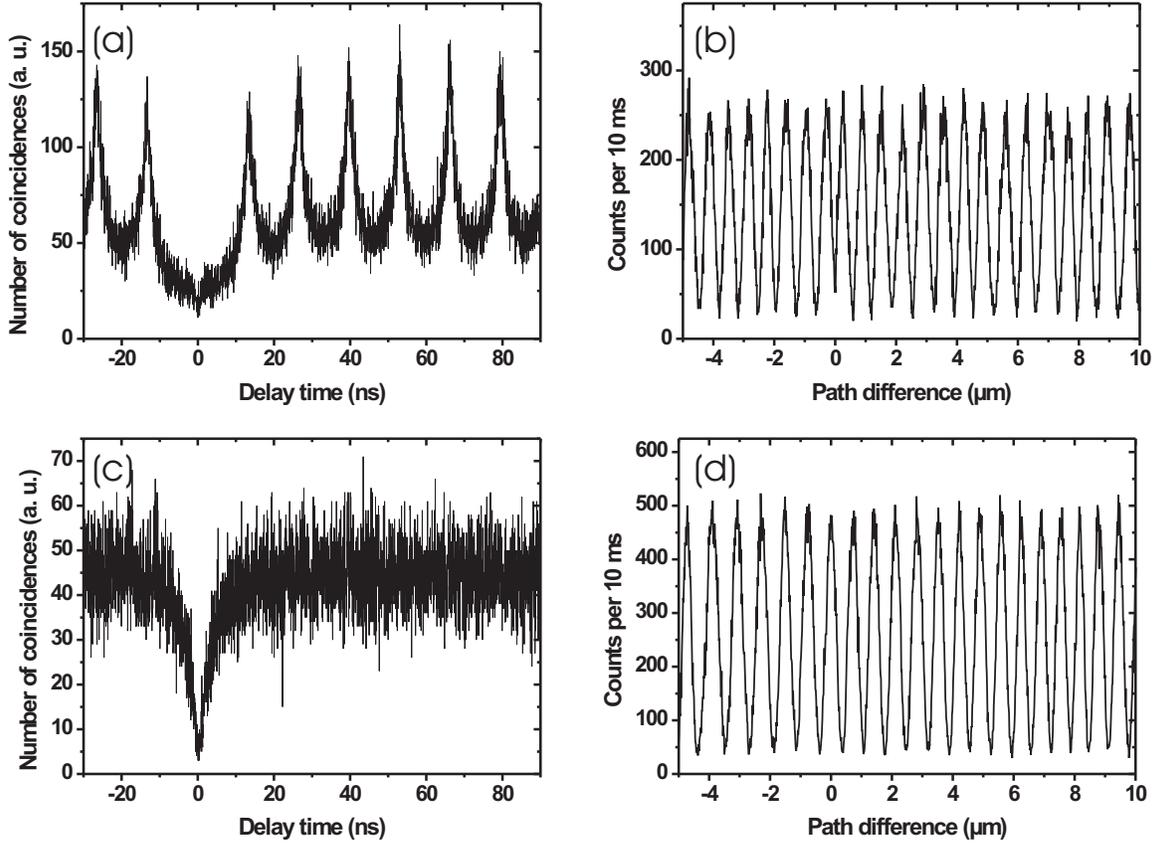}
\caption{\label{Results}(a) and (b) measured correlation function and interference pattern,
respectively, for a single quantum dot with pulsed excitation. (c) and (d) the same, but with
continuous excitation.}
\end{figure}

In our experimental setup (fig. \ref{Setup}) the Michelson interferometer and the Hanbury
Brown-Twiss correlator are combined. Each photon emitted by the quantum dot interferes with itself
at the beam splitter of the Michelson interferometer. When moving one of the interferometer mirrors
forth and back within a few wavelengths, the interference pattern continuously varies between
constructive and destructive interference. Note that in order to determine the visibility of the
interference the intensity from equation \eqref{EqnInt} has to be replaced with the probability
that a photon exits at the fourth port of the interferometer and is not sent back into the
direction of the photon source. The exit of the interferometer directly enters the Hanbury
Brown-Twiss setup where the photons are detected by one of the APDs.

The output pulses of the detectors can be exploited in two ways: First, when looking for
coincidences between pulses from the start and the stop APD, antibunching is observed, revealing
the particle nature of light. The only effect from the change between constructive and destructive
interference on this measurement is an overall change of the coincidence rate, independent of the
delay time between start and stop events. Since the latter is short compared to the time scale of
the arm length variation in the Michelson interferometer, the non normalized second-order coherence
function just changes by a constant factor. Second, one can count the detection pulses of only one
APD, using pulse counting electronics that determines the count rate at a given time. In this case,
the temporal interference pattern, a wave-feature of light, will be observable directly. Since the
detector produces a classical electrical pulse that can, after detection of a photon, be easily
split into two parts, it is also possible to perform these two measurements simultaneously.

Figure \ref{Results} displays the results of such a combined measurement when exciting a quantum
dot with a pulsed ((a) and (b)) and a continuous laser ((c) and (d)), respectively. In these
measurements the piezo, controlling one Michelson arm, was driven by a triangular voltage signal
corresponding to an amplitude of 20 $\mu$m (approx. 29 wavelengths) in the interferometer path
difference, at a modulation rate of 10 mHz. In Figs. \ref{Results}(a) and (c), the autocorrelation
functions of the two measurements are plotted, expressed by the number of coincidences, integrated
over 2 hours (pulsed) and 1 hour (cw). The nonclassical antibunching effect is clearly visible
since the number of coincidences exhibits a pronounced minimum at zero time delay. In contrast,
Figs. \ref{Results}(b) and (c) depict the single-detector count rate, at an integration time of 10
ms, in dependence of the path difference in the interferometer, showing the expected first-order
interference pattern that reveals the wave-like nature of the emitted single-photon radiation.

The fact that the number of coincidences does not completely drop to zero, as theoretically
expected for our single-photon fields, has various reasons. First, the limited time resolution of
the detectors causes the measured signal to wash out and to slightly lift the minimum. Second, an
incoherent spectral background results in the transmission of photons from other transitions
through the bandpass filter and influences the photon statistics. Moreover, a re-excitation process
is observed in this type of quantum dots \cite{NJP}. This originates from the trapping of free
charge carriers inside the dot. In this way re-excitation of the quantum dot and emission of a
second, stopping photon can occur even before the next laser excitation pulses, which especially
affects the pulsed measurement.

\section{Conclusions}

The described combination of the two experimental techniques, Michelson interferometer and Hanbury
Brown-Twiss correlation setup, forms an extension to the experiments of Grangier et al.
\cite{Grangier} and Braig et al. \cite{KurtsieferWaveParticle}, as one and the same photon
contributes to both the measured interference pattern and the antibunched correlation function. In
this sense the described experiment is similar to the classic experiment of Taylor \cite{Taylor}
and to the experiments described in \cite{Janossy,ReynoldsSpartalianScarl}, but gives an
unequivocal evidence of the particle nature of light: Instead of using weak light fields with
classical photon number statistics (in this case a super-Poissonian photon number distribution),
the antibunching effect shows that the quantum dot photoluminescence represents photon number
states, that can only be described within the frame of quantum mechanics.

Our experiments illustrate that single-photon sources are very useful tools not only in quantum
information processing, but also in demonstrations of the fundamental principles of quantum
mechanics. So far, such demonstrations have been dominated by experiments based on photons from
single atoms \cite{Kuhn} and ions \cite{Maurer} or from parametric down conversion
\cite{Zeilinger4Photon}. While these systems still have the advantage of lower decoherence, systems
based on solid state devices have the great potential of integrability and scalability. Thus, the
generation of single photons is a first step towards deterministic number states or entangled
states using semiconductor quantum dots.

\begin{theacknowledgments}
We would like to thank V. Zwiller for fruitful discussion. We gratefully thank W. Seifert for
providing the sample. This work was supported by the Deutsche Forschungsgemeinschaft DFG, grant
BE2224/1, Sonderforschungsbereich (SFB) 296, the Land Berlin, and the European Union (EFRE).
\end{theacknowledgments}

\bibliographystyle{aipproc}

\bibliography{wp2_arxiv}

\end{document}